\documentclass[doublecol]{epl2}
\usepackage{graphicx,color}
\usepackage {ulem}

\title{Interface driven magnetoelectric effects in granular CrO${_2}$}

\author{A. Bajpai\inst{1} \and P. Borisov \inst{2} \and S. Gorantla \inst{1} \and R. Klingeler \inst{3,1} \and J. Thomas\inst{1} \and T. Gemming \inst{1}\and W. Kleemann \inst{2}\and B. B\"{u}chner \inst{1}}

\shortauthor{Bajpai \etal}

\institute{
  \inst{1} Leibniz-Institute for Solid State and Materials Research, IFW Dresden, 01171 Dresden, Germany\\
  \inst{2} Angewandte Physik, Universit\"{a}t Duisburg-Essen, 47048 Duisburg, Germany\\
  \inst{3} Kirchhoff-Institute for Physics, Universit\"{a}t Heidelberg, 69120 Heidelberg, Germany\\ }

\abstract{Antiferromagnetic and magnetoelectric Cr${_2}$O${_3}$-surfaces strongly affect the electronic properties
in half metallic CrO${_2}$. We show the presence of a Cr${_2}$O${_3}$ surface layer on CrO${_2}$ grains by high-resolution transmission electron microscopy. The effect of these surface layers is demonstrated by measurements
of the temperature variation of the magnetoelectric susceptibility. A major observation is a sign
change at about 100 K followed by a monotonic rise as a function of temperature. These electric
field induced moments in CrO${_2}$ are correlated with the magnetoelectric susceptibility of pure
Cr${_2}$O${_3}$. This study indicates that it is important to take into account the magnetoelectric character of thin surface layers of Cr${_2}$O${_3}$ in granular CrO${_2}$ for better understanding the transport mechanism in this system. The observation of a finite magnetoelectric susceptibility near room temperature may find utility in device applications. }

\begin{document}

\maketitle

\section{Introduction}
 Advances in the field of spintronics are driven by the identification of materials with high degree of spin polarization
(such as half metalic ferromagnets) and  materials in which magnetism can be tuned not only with
an applied magnetic field but also through an applied electric field ( magnetoelectric materials)
\cite{Cheong1,Ramesh,Fiebig1,Dorr}.  This in turn has lead to the study of composites of
ferromagnetic and ferroelectric materials, which can exhibit properties superior to a single-phase
compound \cite{Fiebig1}.

 A potential candidate for spintronic applications is CrO${_2}$, which is  a well established half metallic ferromagnet ($T_{\rm C}\approx 393$\,K) \cite{Cheong2,Groot,Soulen,Parker}. Owing to the large spin polarization value as well as it being a room temperature ferromagnet, CrO${_2}$ is widely explored, both in thin films \cite{Soulen,Parker,Gupta1,Ivanov1,Rudiger,Ivanov2} and in bulk \cite{Coey1,Dai1,Dai2} form. A major limiting factor in realization of actual
devices based on CrO${_2}$ has been the issue of stability of its surface\cite{Gupta1,
Ivanov1,Rudiger,Ivanov2,Coey1,Dai1,Dai2}. For instance, it easily converts to other stable oxides,
such as Cr${_2}$O${_3}$. However, in case of granular CrO${_2}$, it was recognised that this could
be a rather advantageuos situation as far as magnetoresistive properties are concerned. The
presence of Cr${_2}$O${_3}$ as an insulating surface layer  makes granular CrO${_2}$ a natural
source of a Magnetic Tunnel Junction \cite{Cheong2,Coey1,Dai1}. This, together with the fact that
Cr$_{2}$O$_{3}$ as the insulating surface is  both antiferromagnetic (AFM) and magnetoelectric
(ME) \cite{Rado,Astrov,Folen,Hornreich,Shtrikman,Martin,Brown} below $T_{\rm N}\approx 307$\,K in
bulk, makes granular CrO${_2}$ interesting from both the fundamental as well as the technological
point of view.  Correspondingly, this system provides the opportunity of a high spin polarization
intrinsically associated with CrO${_2}$ as well as of a significant magnetoelectricity arising
from the Cr${_2}$O${_3}$ surfaces.

 It is emphasized that while the phenomenon of magnetoelectricity is well established for Cr${_2}$O${_3}$
in the bulk single crystalline and polycrystalline forms \cite{Rado,Astrov,Shtrikman}, these features have not been explored when Cr${_2}$O${_3}$ appear as a grain boundary in granular CrO${_2}$. In this work we first
confirm the presence of the crystalline  surface layer of Cr${_2}$O${_3}$ by TEM investigations.
Further, we  present direct experimental measurement of magnetoelectric susceptibility
in granular CrO${_2}$ in which Cr${_2}$O${_3}$ appears as a surface layer outside the
CrO${_2}$ grain.

\section{Experimental Details}
  The sample is in the form of loosely sintered pellets of CrO${_2}$, containing  grains in the form of  micron  size
rods.  The synthesis and other characterization details can be found elsewhere \cite{Bajpai1}. It
is important to note that these samples have shown much better magnetoresistive properties than
commercial powders \cite{Bajpai2}. The transmission electron microscopic (TEM)
investigation on individual CrO${_2}$ grains (derived from the  same pellet) was performed using a FEI Tecnai F30 S-Twin.  The measurements of the ME susceptibility have been performed  using a superconducting quantum interference device (SQUID), MPMS-5S from Quantum Design, by modification of its ac susceptibility option \cite{Pavel}. The electrical conductivity has been measured in the standard four probe geometry, using a Quantum Design PPMS.

\section{Results and discussions}

\subsection{Transmission Electron Microscopy Measurements}

For the TEM investigations, a single grain with a length of the order of a few $\mu$m and
thickness of the order of 500 nm was isolated from the same pellet (Fig.1a) on which the ME
susceptibility  and  resistivity  measurements have been conducted. This sample shows a saturation
magnetization of the order of 120 emu/g \cite{Bajpai1}. The TEM specimen is prepared by pressing a
sparse layer of powder onto the grid and identifying isolated grains from the edges (Fig.1b).
From the relatively thick grains a suitably thin edge was used for high resolution
transmission electron microscopy (HRTEM). Fig.1b shows an individual CrO${_2}$ grain, the arrowhead in the figure represents the region from which the HRTEM image is obtained. The CrO${_2}$ grain was tilted to the $\left[-111\right]$ zone axis using Kikuchi lines prior to HRTEM imaging which is shown in Fig.1c. The Fast Fourier
Transform (FFT) diffractogram from the highlighted region in Fig.1c was analyzed using the ELDISCA
software \cite{Thomas}, for determining the crystal structure. The surface layer is found to be
polycrystalline Cr${_2}$O${_3}$, while no possible zone axis was found when a FFT diffractogram
was analyzed with CrO${_2}$ crystal structure.  The presence of similar continuous Cr${_2}$O${_3}$
layers was observed along the edges of other CrO${_2}$ grains.

\begin{figure}
\onefigure[width=1.0\columnwidth,clip]{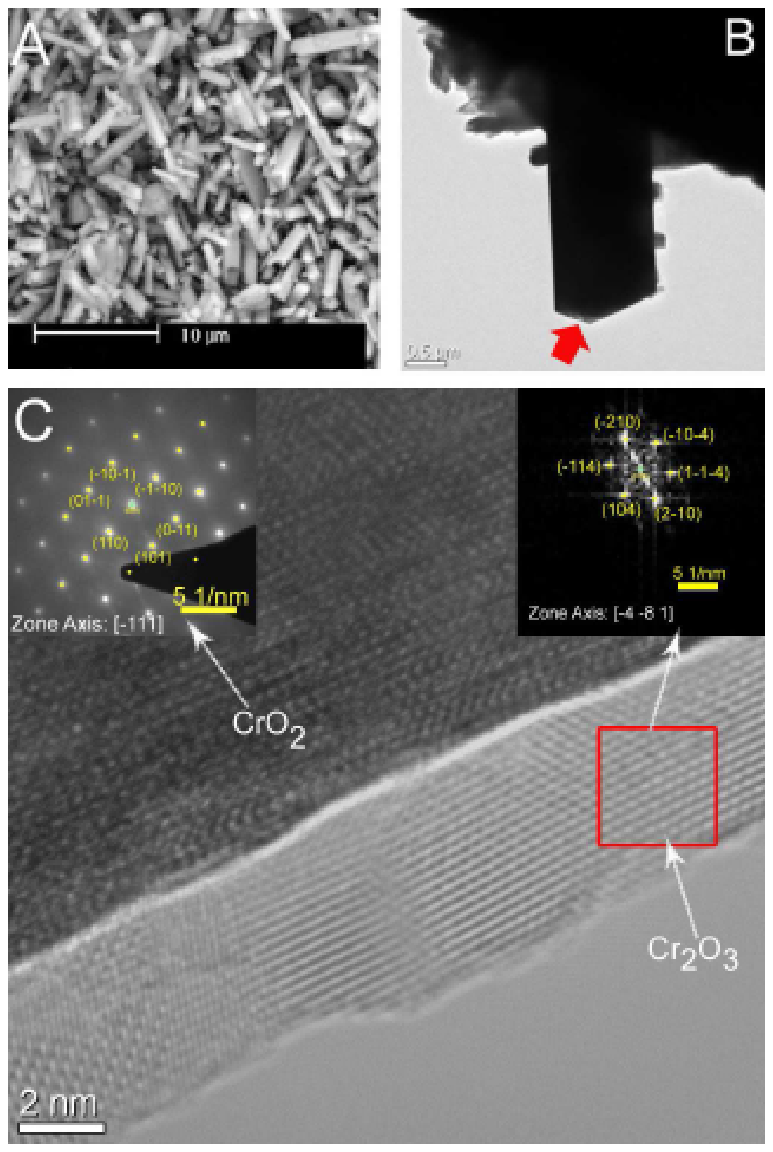} \caption{(a) SEM image of a
pellet containing the CrO${_2}$ grains in the shape of micron size rods.  The bright field TEM
image of an individual CrO${_2}$ grain is shown in (b). The HRTEM image of CrO${_2}$ grain with
Cr${_2}$O${_3}$ oxide layer is shown in (c).  The right inset in (c) shows the FFT diffractogram,
extracted from the region represented by the rectangular box in (c), overlaid by the calculated
diffraction pattern for the [-4-8 1] zone axis of Cr${_2}$O${_3}$. The left inset shows the
experimental diffraction pattern of CrO${_2}$ grain overlaid by a calculated pattern for the
[-111] zone axis.} \label{fig.1}
\end{figure}

The presence of Cr${_2}$O${_3}$  as a surface layer is consistent with the better
tunneling magnetoresistance as observed in these samples \cite{Bajpai2}. Such a layer of low
crystallinity has  earlier  been  seen in TEM studies on commercial powders \cite{Dai1} and very
recently in nanorods of CrO${_2}$ \cite{Song}, synthesized using our synthesis route \cite{Bajpai1}.
However the direct observation of the crystalline Cr${_2}$O${_3}$ as a surface layer, such as
shown in Fig.1c, is rare.

\subsection{Magnetoelectric susceptibility}

 The linear ME susceptibility  $\alpha$${_i}$${_j}$  is
defined using  $\mu_{0}M{_i}$ $\propto$ $\alpha$${_i}$${_j}$E${_j}$ or P${_j}$ $\propto$
$\alpha$${_i}$${_j}$H${_i}$,where M${_i}$ is the electric-field-induced  magnetization and P${_j}$
is the magnetic-field-induced  polarization \cite{Rado}. In this work, we have measured electric-
field-induced magnetization. Our ME susceptibility measurements are conducted by applying an ac
voltage, $u(t)=u_{0}\cos\omega t$ across the sample  and measuring the induced ac magnetic moment,
$m(t)=m'\cos\omega t +m''\sin\omega t$ in  zero external magnetic field.The schematic of the
measurement, which can be done in both \textit{constant applied voltage amplitude}
($u_{0}=\mathrm{const}$) and \textit{constant applied current amplitude} mode
($i_{0}=\mathrm{const}$), is shown in Figure 2. The details of the measurement procedure can be
found elsewhere \cite{Pavel}. The lower panel of Figure 2 shows a complex plane representation of
the real (m') and the  imaginary (m") part of the ME susceptibility as measured on the CrO${_2}$
sample at $i_{0}=\mathrm{const}$ mode at 10 Hz.

\begin{figure}
\onefigure[width=8cm,height=3cm]{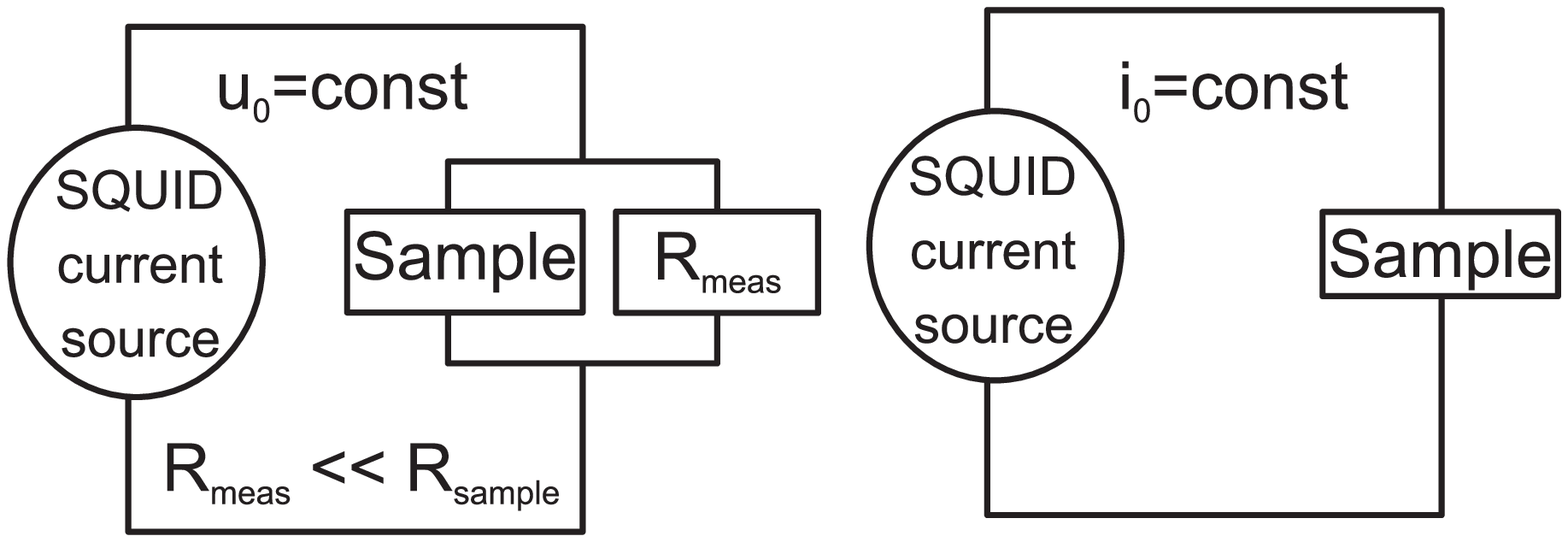}
\onefigure[width=7cm,height=4cm]{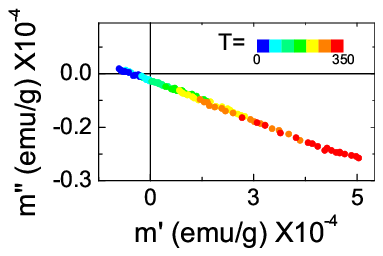}
\caption{ The upper left panel depicts the schematic of the ME susceptibility measurement in constant voltage amplitude mode, while the upper right panel shows the same in constant current amplitude mode. The lower panel displays the complex plane representation ($m''$ vs. $m'$) of the real and the imaginary parts of the electric field induced moment as measured in granular CrO${_2}$ at different temperatures.}
\label{fig.2}
\end{figure}

Fig.3(a)  shows the real part of the electric field induced moment as a function of
temperature for granular CrO${_2}$. The data are obtained in both cooling and heating cycles,
measured in constant current amplitude mode. The moments exhibit a change of sign below 100 K,
followed by a rapid rise in magnitude up to the maximum temperature of  350 K. A substantial
thermal hysteresis is observed in cooling and heating cycles, which becomes pronounced above
200~K.  Fig.3(b) compares the ME susceptibility in a single crystal of Cr${_2}$O${_3}$
( blue stars) with that measured on the granular CrO${_2}$ (red circles). Both measurements were performed in heating cycle using the same experimental set up.
The ac electric-field amplitude for granular CrO${_2}$ and Cr${_2}$O${_3}$ were 1 V/m and 63 kV/m,
respectively. In addition, the single crystal of Cr${_2}$O${_3}$ was ME annealed in the magnetic
field of 6000 Oe and in dc electric field of 300 kV/m \cite{Pavel} whereas no annealing fields
were required for CrO${_2}$. The data on Cr${_2}$O${_3}$ are multiplied by a factor of 10 so as to
compare its temperature variation with that measured on granular CrO${_2}$. This comparison
reveals  that in both cases the ME signal changes sign at about 100 K. However, unlike what is
observed in pure Cr${_2}$O${_3}$, the ME moments in granular CrO${_2}$ do not vanish at the
$T_{\mathrm{N}}$ of Cr${_2}$O${_3}$,  but continue to rise further till the measured temperature
of 350K. Fig.~3c shows that the resultant moment exhibits a linear variation as a function of the applied-
electric-field amplitude measured at a fixed temperature, as is expected for linear ME
effects, typically seen in pure Cr${_2}$O${_3}$.

\begin{figure}
\onefigure[width=8cm,height=12cm]{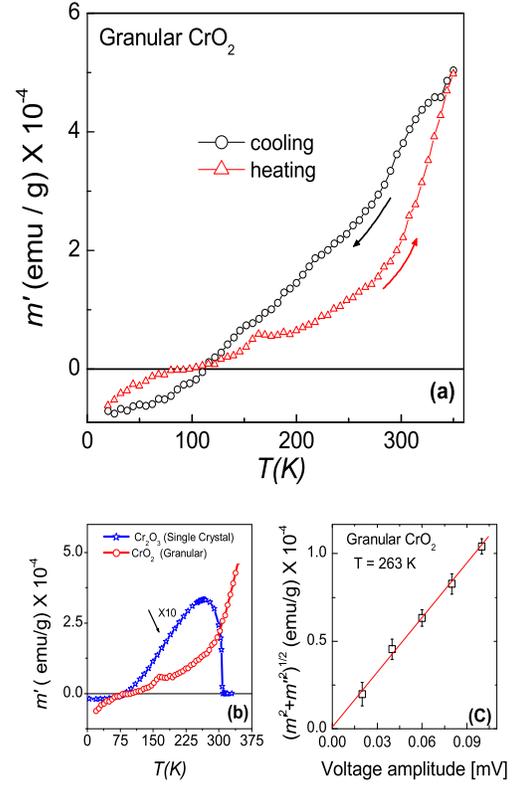}
\caption{ (a) Temperature dependence of the real part of the electric field induced moment ($m'$)
in cooling and heating cycles, showing thermal hysteresis and a change of sign around 100K. (b) $m'$ in a single crystal of Cr${_2}$O${_3}$ (blue stars) and granular CrO${_2}$ (red circles). The data set on  Cr${_2}$O${_3}$ is multiplied by a factor of 10. (c) Resultant amplitude of the moment varying linearly  as a function of the applied electric field amplitude at
263 K. }
\label{fig.3}
\end{figure}

We first focus our attention to the change of sign in ME moments at about 100 K. It is to be noted
that the sign of the induced moments for such measurements is arbitrary, but is fixed for a
particular connection during a run. Therefore, the change of sign during one set of run is not
arbitrary. It is recalled that the ME susceptibility measured on pure Cr${_2}$O${_3}$ shows a
similar change of sign at around 100 K, which is well documented in literature \cite{Rado,
Hornreich,Pavel}. This feature in pure Cr${_2}$O${_3}$ is understood to arise from an interplay
between a two-ion term (electrically induced change in the exchange interactions in sublattices
that disturbs the mutual compensation of antiparallel spins) and a single-ion term (due to the
electric field induced g-shift) which have opposite signs. Although Cr${_2}$O${_3}$ appears only
as grain boundary in granular CrO${_2}$, both of them, granular CrO${_2}$ and pure Cr${_2}$O${_3}$
exhibit a change of sign in the ME susceptibility, occurring in a similar temperature range and
this feature is indicatative of strong interface effects. This provides compelling evidence that
the ME response  in granular CrO${_2}$ is modulated by features intrinsic to the pure
Cr${_2}$O${_3}$ phase.

We also observe that the ME signal is much larger in magnitude in relatively small applied
electric field in CrO$_{2}$ and no magnetoelectric annealing is required as in case of
Cr${_2}$O${_3}$. A rough attempt to extract the size of the
corresponding magnetoelectric coupling for CrO${_2}$ yields $\alpha$ ($\approx
\mu_{0}M{_z}$/E${_z}$) in the order of $10^{-7}$ s/m.  This estimation is based on the experimentally observed
electric field induced moment (m') and secondary effects such as the
anisotropy of the ME effect, AFM domains and the polycrystalline nature of our sample have been
neglected. This value is about 5 orders of magnitude larger
than the maximal value of $\alpha$ for the pure Cr${_2}$O${_3}$ which exhibits $\alpha_{zz}\approx 4$
ps/m \cite{Astrov,Rado,Pavel}. However, we emphasize that any comparison of $\alpha$
e.g. for the two samples studied at hand, should be taken very cautiously.

In this context, it is important to recall that while Cr${_2}$O${_3}$ is insulating, granular
CrO${_2}$ shows activated transport with relatively low resistivity values of the order of a few
$\Omega$cm  \cite{Bajpai2,Bajpai3}. This may lead to additional contributions to the ME effect and
need to be distinguished from the classical ME effects intrinsic to pure Cr${_2}$O${_3}$. For
instance, the low resistivity of our samples  may lead to generation of moments induced due to the
surface currents, which we refer to as the Oersted moment, to differentiate it from ME response.
During a ME measurement at a constant current amplitude, this contribution should be independent
of the temperature - if we ignore the modulations driven by the ferromagnetic interactions.

 In order to investigate the possible effect of Oersted moments, we conducted an experiment in
the same configuration on a graphite sample. This sample consists of thin graphite
sheets, commercially available from Goodfellow and three such sheets were glued together by silver
paste so that the resistivity of this sample is of the order of a few $\Omega$cm, similar to the
CrO${_2}$ sample. The sample geometry was similar to  the CrO${_2}$ sample, too. In
Fig.~4, we compare the normalized electric field induced moments as observed in a cooling cycle
for both the granular CrO${_2}$ ($i_{0}=125$ $\mu$A$ =\mathrm{const}$) and the graphite sample
($i_{0}=175$ $\mu$A$ =\mathrm{const}$). The data indeed imply the presence of finite
Oersted moments in the graphite model sample. However,  the observed moments in the graphite
sample show no change of sign and they are temperature independent. This is unlike what
is observed in granular CrO${_2}$. Thus, although Oersted moments may affect the
magnitude of the observed signal, these contributions cannot explain the strong temperature
dependence of the electric-field-induced moments as is seen in granular CrO${_2}$.

 We also note that it is nontrivial to estimate the magnitude of surface currents  in  our sample with randomly orientated grains with insulating grain boundary. Though we have tried to estimate the maximal possible magnitude of ME moments, taking into account the  Cr${_2}$O${_3}$ surface layer outside the grains \footnote{Under assumptions that the CrO$_{2}$ sample consists of spherical grains with average diameter of 5$\mu$m and with grain boundaries layer of 2 nm Cr$_{2}$O$_{3}$, the electric field drop at the grain interface is about $10^{3}$  times  the external electric field. As the  volume ratio of Cr$_{2}$O$_{3}$ outside a single grain is about $10^{-3}$, the total electrically induced moment is finally the same as in the case of 100\% Cr$_{2}$O$_{3}$. }. These calculations show that the magnitude of the ME moments is still  4 to 5 orders of magnitude smaller than what is observed experimentally  and therefore the exact scenario cannot be understood unless interface effects are taken into accout.

We also plot the normalized conductance measured at a fixed bias current of 100 $\mu$A
for granular CrO${_2}$ in the same graph (Fig.~4). This comparison reveals that the
functional form of the observed moments in granular CrO${_2}$ is strikingly different from what is
expected from the surface currents  associated to the Oersted moment. The observation of
change of sign was also reconfirmed by measurements repeated in both  constant current amplitude
(Fig.~4, $i_{0}=125$ $\mu$A) and constant-voltage-amplitude mode (not shown).  These experimental observations  strongly
indicate that the origin of electric-field-incuced moments in our sample is correlated with ME
effects in Cr${_2}$O${_3}$.

    While change of sign as well as linear variation of moments as a
function of electric field (Figure 3c) point towards the modulation driven by AFM/ME
Cr${_2}$O${_3}$, the continuous increase of the moment as a function of temperature right up to
350 K is puzzling. This is contrary to what is seen in the pure Cr${_2}$O${_3}$ phase where the ME
moment vanishes at $T_{\rm{N}}$ of Cr${_2}$O${_3}$ (307\,K). Though such effects are not uncommon
and plenty of work - particularly in  tailor made FM/AFM systems - reports  unusual exchange bias
effects that exist beyond the N\'eel temperature of its AFM component\cite{Sahoo,Shi,Zaag}. These
features are usually attributed to strain or proximity effect and we also have seen some
manifestations in our samples\cite{Bajpai4}. However, the exact physical mechanisms that leads to
such observation are not well explored.

\begin{figure}
\onefigure[width=8cm,height=8cm]{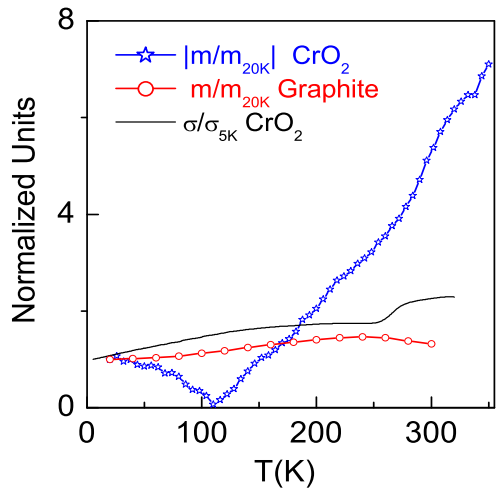}
\caption{ Comparison of normalized values of electric field
induced moment recorded in cooling cycle for CrO${_2}$ sample (blue stars) and a graphite sheet
(red circles) having resistivity in similar range with that of CrO${_2}$ sample. The black line is
the normalized conductance of CrO${_2}$ measured in the bias current of 100 $\mu$A.}
\label{fig.4}
\end{figure}

 In this context, it is  recalled that in general, the microscopic origin of ME effects is
understood to arise from four basic mechanisms such as g tensor, single ion anisotropy, symmetric
and antisymmetric exchange mechanism\cite{Gehring1,Gehring2}. Depending on the strength of
interaction, the total ME effect is governed by one of these mechanisms or their combinations. For
instance, the two-ion contribution, related to  a net change in the bulk magnetization of AFM ,
arising due to the inequivalence of the two sublattices on application of the electric field,
vanishes above the N\'eel temperature.  However, the ME annealing effects as observed in pure
Cr${_2}$O${_3}$ are explained in terms of the atomic single-ion ME coefficients which exist in the
paramagnetic regime (i.e. above 307 K) in Cr${_2}$O${_3}$ but this effect is enhanced when
long-range AFM order sets in\cite{Martin}. The usual ME effect (related to the AFM order) is
itself a weak effect and to observe the change due to individual moments is even more challenging.
However, it is in principle possible to observe the ME effect related to the change in the
magnitude of an individual moment on application of the electric field\cite{Gehring 1,Gehring2}.

 In our case, the Cr${_2}$O${_3}$ appears at the
interface of another long-range ordered magnet (CrO${_2}$), whose $T_{\rm{C}}$ is about
400\,K. In this situation, the ME moments arising due to  single-ion ME coefficients of  Cr${_2}$O${_3}$  may  still be finite above $T_{\rm{N}}$, due to proximity effects \cite{Song,Cheng}. Though we could
measure the ME moments up to about 350\,K, we speculate that they may eventually diminish only
above 400K. Prior experimental data on thin films of CrO${_2}$
having a surface layer of Cr${_2}$O${_3}$ as well as bilayers of  CrO${_2}$/Cr${_2}$O${_3}$ have
also shown some interesting interface effects \cite{Cheng,Frey}.

This study indicates that it is important to explore the influence of the AFM/ME
character of Cr${_2}$O${_3}$ when it appears at the interface, particularly in transport
measurements. In these samples, we do see unusual thermal history and bias current effects in
resistivity measurements (unpublished). These data also point towards the need to take into
account the ME character of the grain boundary in order to understnd the transport mechanism in
this system.

In conclusion, from direct measurement of high-resolution transmission electron microscopy we show
that our CrO${_2}$ grains contain  a surface layer of \textit{polycrystalline}
Cr${_2}$O${_3}$. This can account for the much improved tunneling magnetoresistance as has earlier
been observed in these samples.  The influence of the magnetoelectric character of the surface
layer is visible in magnetoelectric susceptibility measurements. The observation of finite electric-field-induced moments right up to
350K  in granular CrO${_2}$ can  be understood to arise from interface effects, taking into account the modulations
driven by the ME character of Cr${_2}$O${_3}$.  A finite ME susceptibility at room temperature
may prove to be useful in CrO${_2}$ based devices.

\acknowledgments
     A.B. acknowledges support through a EU Marie Curie IIF fellowship under project 040127-NEWMATCR. P.B. acknowledges support by the DFG through SFB 491.
Part of the work was funded by the DFG (HE 3439/6)


\begin{thebibliography}{}

\bibitem{Cheong1}
\Name{S-W. Cheong  \and M. Mostovoy}
\REVIEW{Nat. Mater.}{6}{2007}{13}.

\bibitem{Ramesh}
\Name{R. Ramesh \and N. A. Spaldin }
\REVIEW{Nat. Mater.}{6}{2007}{21}.

\bibitem{Fiebig1}
\Name{ M. Fiebig}
\REVIEW{J. Phys. D: Appl.Phys.}{38}{2005}{R123}.

\bibitem{Dorr}
\Name{ K. D\"{o}rr }
\REVIEW{J. Phys. D: Appl.Phys.}{39}{2006}{R125}.

\bibitem{Cheong2}
\Name{ H.Y. Hwang \and S-W. Cheong}
\REVIEW{Science}{278}{1997}{1607}.

\bibitem{Groot}
\Name{R.A. Groot \etal}
\REVIEW{Phys. Rev. Lett.}{50}{1983}{2024}.

\bibitem{Soulen}
\Name{R.J. Soulen \etal}
\REVIEW{Science}{282}{1998}{85}.

\bibitem{Parker}
\Name{J. S. Parker, S. M. Watts, P. G. Ivanov \and P. Xiong}
\REVIEW{Phys. Rev. Lett.}{88}{2001}{196601}.

\bibitem{Gupta1}
\Name{ L. Spinu, H. Srikanth, A.Gupta, X.W. Li and Gang Xiao}
\REVIEW{Phy. Rev. B}{62}{2000}{8931}.

\bibitem{Ivanov1}
\Name{ P.G. Ivanov, S.M. Watts and D.M.Lind}
\REVIEW{J.Appl. Phys.}{89}{2001}{1035}.

\bibitem{Rudiger}
\Name{ U. Ruediger \etal,}
\REVIEW{J.Appl. Phys.}{89}{2001}{7699}.

\bibitem{Ivanov2}
\Name{P.G. Ivanov and K.M. Bussmann}
\REVIEW{J. Appl. Phys.}{105}{2007}{07B107}.

\bibitem{Coey1}
\Name{J.M.D Coey \etal }
\REVIEW{Phys. Rev. Lett.}{80}{1998}{3815}.

\bibitem{Dai1}
\Name{J. Dai \etal}
\REVIEW{Appl. Phys. Lett.}{77}(2000){2840}.

\bibitem{Dai2}
\Name{ J. Dai  \and J. Tang}
\REVIEW{Appl. Phys. Lett}{63}{2001}{064410}.


\bibitem{Rado}
\Name{G.T. Rado \and V.J. Folen}
\REVIEW{Phys. Rev. Lett.}{7}{1961}{310}.

\bibitem{Astrov}
\Name{D.N.Astrov}
\REVIEW{Sov. Phys. JETP}{11}{1960}{708}.


\bibitem{Folen}
\Name{ V.J. Folen, G.T. Rado \and E.W. Stalder.}
\REVIEW{Phys. Rev. Lett.}{6}{1961}{607}.

\bibitem{Hornreich}
\Name{R. Hornreich \and S.Shtrikman}
\REVIEW{Phys. Rev.}{161}{1961}{506}.

\bibitem{Shtrikman}
\Name{S. Shtrikman \and D. Treves}
\REVIEW{Phys. Rev.}{130}{1963}{986}.

\bibitem{Martin}
\Name{ T.J. Martin and J.C. Anderson}
\REVIEW{Phys. Lett.}{11}{1964}{109}.

\bibitem{Brown}
\Name{P.J. Brown \etal}
\REVIEW{J. Phys.: Cond. Matter}{10}{1998}{663}.

\bibitem{Pavel}
\Name{P. Borisov \etal}
\REVIEW{Rev. Sci. Instrum.}{78}{2007}{106105}.

\bibitem{Bajpai1}
\Name{A Bajpai  \and A.K. Nigam }
\REVIEW {US Patent }{7276226}{
}.
\bibitem{Bajpai2}
\Name{A. Bajpai  \and  A.K. Nigam }
\REVIEW{Phys. Rev. B}{75}{2007}{064403}.

\bibitem{Bajpai3}
\Name{ A. Bajpai  \and A.K. Nigam}
\REVIEW{J. Appl. Phys.}{101}{2007}{10391}.

\bibitem{Thomas}
\Name{J. Thomas and T. Gemming}
\REVIEW{Springer, ISBN 978-3-540-85301-5} {2008}{231}.

\bibitem{Song}
\Name{Y. Song, A.L. Schmitt and Song Jin }
\REVIEW{Nano Lett.}{8}{2008}{2356}.

\bibitem{Cheng}
\Name{ Cheng R. \etal}
\REVIEW{Phys. Lett. }{A302}{2008}{21}.

\bibitem{Frey}
\Name{N.A. Frey  \etal}
\REVIEW{Phys. Rev. B }{74}{2006}{024420}.

\bibitem{Sahoo}
\Name{S. Sahoo\etal}
\REVIEW{Appl. Phys. Lett.}{91}{2007}{172506}.

\bibitem{Shi}
\Name{H. Shi  \etal}
\REVIEW{Phys. Rev. B}{69}{2004} {214416}.

\bibitem{Zaag}
\Name{P.J. Van der Zaag \etal }
\REVIEW{Phys. Rev. Lett}{84}{2000}{6102}.

\bibitem{Bajpai4}
\Name{A. Bajpai \etal }
\REVIEW{J. Phys: Condens Matter} {22}{2010}{096005}.

\bibitem{Gehring1}
\Name{O.F. de Alcantara Bonfim  \and G.A. Gehring}
\REVIEW{Adv. Phys.} {29}{1980} {731}.

\bibitem{Gehring2}
\Name{G.A. Gehring} \REVIEW{Ferroelectrics} {161} {1993} {275}.



\end{thebibliography}
\end{document}